\documentclass[prb,aps,english,twocolumn,superscriptaddress]{revtex4-1}
\usepackage[Symbol]{upgreek}
\usepackage{amssymb}
\usepackage{amsmath}
\usepackage{graphicx}
\usepackage[colorlinks=true,linkcolor=black, citecolor=blue, urlcolor=blue,
	unicode=true]{hyperref}


\begin{document}
\preprint{}

\newcommand{\spinel}[3]{$\textrm{#1}\textrm{#2}_2\textrm{#3}_4$}
\newcommand{\spinelions}[3]{$\textrm{#1}^{2+}\textrm{#2}^{3+}_2\textrm{#3}^{2-}_4$}
\newcommand{\mvo}{\spinel{Mg}{V}{O}}
\newcommand{\angs}{\mbox{\normalfont\AA}}

\title{Transverse acoustic phonon anomalies at intermediate wavevectors in MgV$_{2}$O$_{4}$}


\newcommand{\tum}{Physik-Department E21, Technische Universit\"at M\"unchen (TUM),
    James-Franck-Str. 1, 85748 Garching, Germany}
\newcommand{\mlz}{Heinz-Maier-Leibnitz-Zentrum (MLZ), Technische Universit\"at
    M\"unchen (TUM), Lichtenbergstr. 1, 85747 Garching, Germany}
\newcommand{\lns}{Laboratory for Neutron Scattering and Imaging (LNS), Paul
    Scherrer Institute (PSI), 5232 Villigen PSI, Switzerland}
\newcommand{\edinburgh}{School of Physics and Astronomy and Centre for Science
    at Extreme Conditions, University of Edinburgh, Edinburgh EH9 3FD, United Kingdom}
\newcommand{\juelichill}{J\"ulich Centre for Neutron Science (JCNS),
    Forschungszentrum J\"ulich GmbH, Outstation at Institut Laue-Langevin,
    Bo\^ite Postale 156, 38042 Grenoble Cedex 9, France}
\newcommand{\mpifkf}{Max-Planck-Institut f\"ur Festk\"orperforschung,
    Heisenbergstr. 1, 70569 Stuttgart, Germany}
\newcommand{\spsms}{D\'epartement de Recherche Fondamentale sur la Mati\`ere
    Condens\'ee, SPSMS/MDN, CEA Grenoble, 38054 Grenoble, France}
\newcommand{\ral}{ISIS Pulsed Neutron and Muon Source, STFC Rutherford Appleton
    Laboratory (RAL), Harwell Campus, Didcot, Oxon, OX11 0QX, United Kingdom}
\newcommand{\nanotech}{London Centre for Nanotechnology and UCL Centre for Materials
    Discovery, University College London, 17-19 Gordon Street, London WC1H 0AH.
    United Kingdom}

\author{T. Weber}
\email{tobias.weber@tum.de}
\affiliation{\tum}
\affiliation{\mlz}

\author{B. Roessli}
\affiliation{\lns}

\author{C. Stock}
\affiliation{\edinburgh}

\author{T. Keller}
\affiliation{\mpifkf}
\affiliation{\mlz}

\author{K. Schmalzl}
\affiliation{\juelichill}

\author{F. Bourdarot}
\affiliation{\spsms}

\author{R. Georgii}
\affiliation{\mlz}

\author{R. A. Ewings}
\affiliation{\ral}

\author{R. S. Perry}
\affiliation{\ral}
\affiliation{\nanotech}

\author{P. B\"oni}
\affiliation{\tum}

\date{\today}

\begin{abstract}
Magnetic spinels (with chemical formula $AX_{2}$O$_{4}$, with $X$ a 3$d$ transition
metal ion) that also have an orbital degeneracy are Jahn-Teller active and hence possess
a coupling between spin and lattice degrees of freedom.
At high temperatures, MgV$_{2}$O$_{4}$  is a cubic spinel based on V$^{3+}$ ions with
a spin $S$=1 and a triply degenerate orbital ground state.
A structural transition occurs at T$_{OO}$=63 K to an orbitally ordered phase with
a tetragonal unit cell followed by an antiferromagnetic transition of T$_{N}$=42 K
on cooling.   We apply neutron spectroscopy in single crystals of MgV$_{2}$O$_{4}$
to show an anomaly for intermediate wavevectors at T$_{OO}$ associated with the
acoustic phonon sensitive to the shear elastic modulus $\left(C_{11}-C_{12}\right)/2$.
On warming,  the shear mode softens for momentum transfers near close to half the
Brillouin zone boundary, but recovers near the zone centre.
High resolution spin-echo measurements further illustrate a temporal broadening with
increased temperature over this intermediate range of wavevectors, indicative of a
reduction in phonon lifetime.  A subtle shift in phonon frequencies over the same
range of momentum transfers is observed with magnetic fields.
We discuss this acoustic anomaly in context of coupling to orbital and charge
fluctuations.

\vspace{0.2cm}
\noindent This is a pre-print of our paper at \url{https://link.aps.org/doi/10.1103/PhysRevB.96.184301},
\copyright{} 2017 American Physical Society.
\end{abstract}

\maketitle

\section{ \label{sec:intro} Introduction }
\begin{figure}[b]
	\includegraphics[width=1\columnwidth]{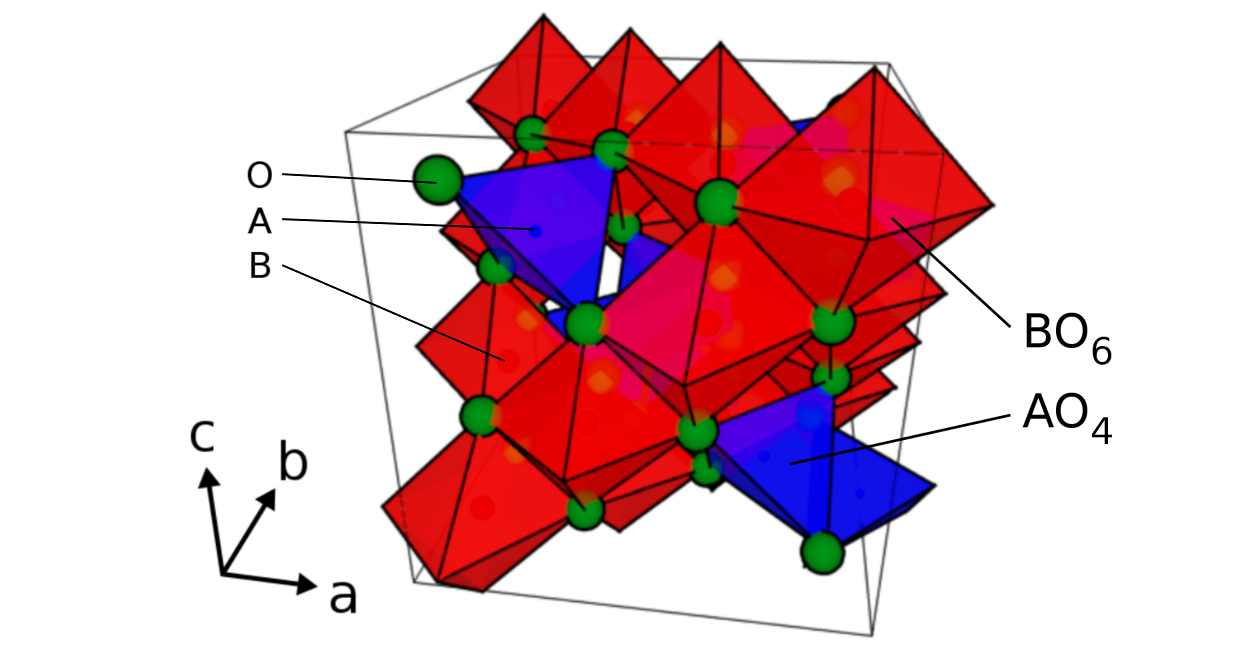}
	\caption{\label{fig:spinel} Conventional unit cell of a spinel. In an
	\spinel{A}{B}{O} spinel, the A metal ions are in a tetrahedral
	$\textrm{AO}_4$ environment that constitutes a diamond lattice, the B metal
	ions form edge-sharing $\textrm{BO}_6$ octahedra arranged in a pyrochlore
	lattice. }
\end{figure}

Magnetically frustrated and orbitally degenerate materials are of high interest
to study the coupling between the lattice, spin, and orbital degrees of
freedom.~\cite{Tokura03,Millis98,Rudolf07}
One important class are the spinels, i.e. minerals possessing the chemical
formula \spinelions{A}{B}{X} where A and B are divalent and trivalent metallic
cations, respectively, and in most cases X are oxygen ions. The anions form a
face-centred cubic lattice with 32 ions in the unit cell. The interstices of the
close-packed structure consist of 8 tetrahedral and 16 octahedral sites which --
in a normal spinel -- are occupied by the smaller A and B cations, respectively
(Fig. \ref{fig:spinel}). The ions occupying the tetrahedral sites constitute a
diamond lattice and the ions in the octahedral sites form a corner-sharing
pyrochlore lattice, which is an archetype for geometrically frustrated magnetic
systems.~\cite{Anderson56}

In spinel vanadates -- \spinelions{A}{V}{O} -- the vanadium ion is situated in a
octahedral environment surrounded by oxygen ions. The crystal field of the oxygen
ions splits the fivefold degenerate $d$ orbitals into $e_g$ and the $t_{2g}$
states with the $e_g$ orbitals lying
higher than the $t_{2g}$ orbitals. The vanadium ions possess two $3d$ electrons
that are distributed over the three degenerate $t_{2g}$ orbitals, namely
$d_{xy}$, $d_{xz}$, and $d_{yz}$.  This distribution of electrons results in one
hole which can occupy any of the three degenerate $t_{2g}$ orbitals resulting
in a three-fold orbital degeneracy.
This orbital occupancy introduces an additional spin-orbit coupling term to
the Hamiltonian.~\cite{McClure59:9,Abragam:book}

Because of the cooperative Jahn-Teller effect~\cite{Kaplan:book,Gehring1975}
for materials with an orbital degeneracy, a structural phase transition
accompanied by orbital ordering~\cite{Radaelli2005,Pandy11} sets in for \mvo{} below a temperature
$T_{OO}$. For $T < T_{OO}$ one of the crystallographic axes -- which is
conventionally taken along $c$ -- compresses with $c/a=0.9941 <1$, lowering the point group symmetry
at the V sites from cubic to tetragonal. The compression along a single axis corresponds to the
$C_{11}-C_{12}$ stability condition of the elastic constants $C_{ij}$ for cubic
crystals~\cite{Cowley76}. Consequently, the triplet degeneracy of the $t_{2g}$ orbital
is partially lifted. One of the vanadium electrons occupies the energetically
lower-lying $d_{xy}$ orbital, while the other electron is shared among the
degenerate $d_{xz}$ and $d_{yz}$ orbitals. The $d_{xy}$ orbitals on the
pyrochlore lattice order in a way that their lobes point to neighbouring
vanadium ions in the $ab$ plane, along $\left[110\right]$ and
$\left[1\overline{1}0\right]$ directions~\cite{Tcherny04}. The occupation of
the $d_{xz}$ and $d_{yz}$ orbitals is an unresolved problem arising from the
competition of three different interaction mechanisms~\cite{Khomskii2014}.
Namely the coupling of the orbital to the elastic strain via the cooperative
Jahn-Teller effect, the Kugel-Khomskii interaction between neighbouring vanadium
ions, and the spin-orbit coupling. Depending on which interaction is assumed to
dominate, different orbital ordering patterns are obtained.

A prominent theoretical model, given by Tsunetsugu and Motome~\cite{Tsunetsugu03},
is based on the Kugel-Khomskii interaction between the
vanadium ions, and results in a ferro-orbital (antiferro-orbital) ordering of
the $d_{xy}$ ($d_{xz}$, $d_{yz}$) orbitals along the [110] ([101], [011])
directions in the pyrochlore lattice.
Because this theory has been found to predict a symmetry for the structural
lattice which is incompatible with measurements~\cite{Wheeler10}, a different
model was devised by O. Tchernyshyov~\cite{Tcherny04}, treating the spin-orbit
interaction to be dominating, followed by the Jahn-Teller, and -- on the weakest
scale -- the Kugel-Khomskii interaction. It predicts an orbital ordering
where one of the electrons on each vanadium ion occupies a complex superposition
of $d_{xz}$ and $d_{yz}$.
For both models~\cite{Tsunetsugu03, Tcherny04}, the spin of the vanadium
electron in the $d_{xy}$ orbital orders antiferromagnetically along chains in
the $\left[110\right]$ and $\left[1\overline{1}0\right]$ directions on the
pyrochlore lattice. In the $bc$ and $ac$ planes, an up-up-down-down spin pattern
forms below the N\'eel temperature $T_N < T_{OO}$. The magnetic moments point
along the $c$ axis of the crystal and a strong (weak) magnetic
coupling is obtained for perpendicular (parallel) chains. In the model by
Tchernyshyov~\cite{Tcherny04}, geometric frustration of the spins in the
perpendicular chains on the pyrochlore lattice gives rise to two (plus two
time-reversed) degenerate ground states.

Neutron inelastic scattering and diffraction experiments by Wheeler
\textit{et al.} \cite{Wheeler10} found a mixture of both of these models, a
real and a complex superposition of the $d_{xz}$ and $d_{yz}$ orbitals, to best
reproduce their experimental data for \mvo{}. The interplay between lattice and
spin/orbital ordering was studied in \mvo{} by Watanabe \textit{et al.}
\cite{Watanabe14} using ultrasound measurements. They found that the $C_{11}$
and $C_{44}$ elastic constants show a softening region for cooling towards
$T_{OO}$ and a large discontinuity at $T_{OO}$, which indeed suggests that there
is a strong coupling between the orbitals of the Jahn-Teller ions and the
lattice strain.  In addition, Watanabe \textit{et al.} \cite{Watanabe14} found a sensitivity of
$C_{11}$ and $C_{44}$ on external magnetic fields up to 7 T in the [110]
direction for the softening region near $T > T_{OO}$ in contrast to
$\left(C_{11}-C_{12}\right)/2$, which does not depend on field.

In this work we present results of inelastic neutron scattering experiments
above and below the orbital-ordering temperature $T_{OO}$ that yield the
dispersion of the acoustic phonon branches in \mvo{} thus complementing
the $q \rightarrow 0$ results of Watanabe \textit{et al.} \cite{Watanabe14}
Unlike ultrasound measurements which probe acoustic fluctuations in the
$lim_{q \rightarrow 0}$ and Raman spectroscopy which is a strictly $q=0$
probe, neutron spectroscopy can investigate excitations associated with
a uniform lattice deformation at all wavevectors, and hence wavelengths,
over the entire Brillouin zone.
We focus our measurements on the acoustic phonons and not the optical
phonons previously reported to exist for energies above $\sim$ 10 meV
using optical techniques.~\cite{Popovic03,Jung08}  We will show that
there exists a range of acoustic phonon wavelengths associated with
the $(C_{11}-C_{12})/2$ elastic constant where the TA mode both softens
in energy and also increases in linewidth indicative of decreased lifetime.

This paper is divided into three main sections including this introduction.
We first describe the neutron scattering experiments studying the low
energy acoustic phonons where the softening in energy over intermediate
wavevectors and then the linewidth broadening indicative of a shortening
of phonon lifetimes is described.
We finally conclude with a discussion comparing our results to theories
for soft modes in Jahn-Teller systems and also a comparison between the
anomalies observed here and in metallic systems with acoustic instabilities.

\section{ \label{sec:exp} Experiments }
\subsection{ \label{sec:char} Sample characterisation }

Three cylindrical single-crystals of \mvo{}, each of about 2 cm height and
0.75 cm diameter were grown using a mirror image furnace.
Characterisation was performed using bulk susceptibility measurements and neutron
diffraction confirming the presence of both a magnetic and structural transition. The bulk
measurements of the heat capacity and the magnetic susceptibility show well
defined structural and antiferromagnetic transitions at temperatures of
$T_{OO} \approx 63 \pm 1 \,\textrm{K}$ and $T_N \approx 42\,\textrm{K}$,
respectively, demonstrating the high quality of the crystals. Using neutron
diffraction on the single-crystals, the space group was confirmed to be
face-centred cubic $\mathrm{Fd\bar{3}m}$ above $T_{OO}$ and tetragonal
$\mathrm{I4_{1}/a}$ below $T_{OO}$ \footnote{Wheeler \textit{et al.}
\cite{Wheeler10} has suggested the space groups of \mvo{} to be
$\mathrm{F\bar{4}3m}$ (cubic) and $\mathrm{I\bar{4}m2}$ (tetragonal) based
on the measurement of weak Bragg peaks which are absent in the
space group $\mathrm{Fd\bar{3}m}$}.

\begin{figure}[htb]
	\includegraphics[width=1\columnwidth]{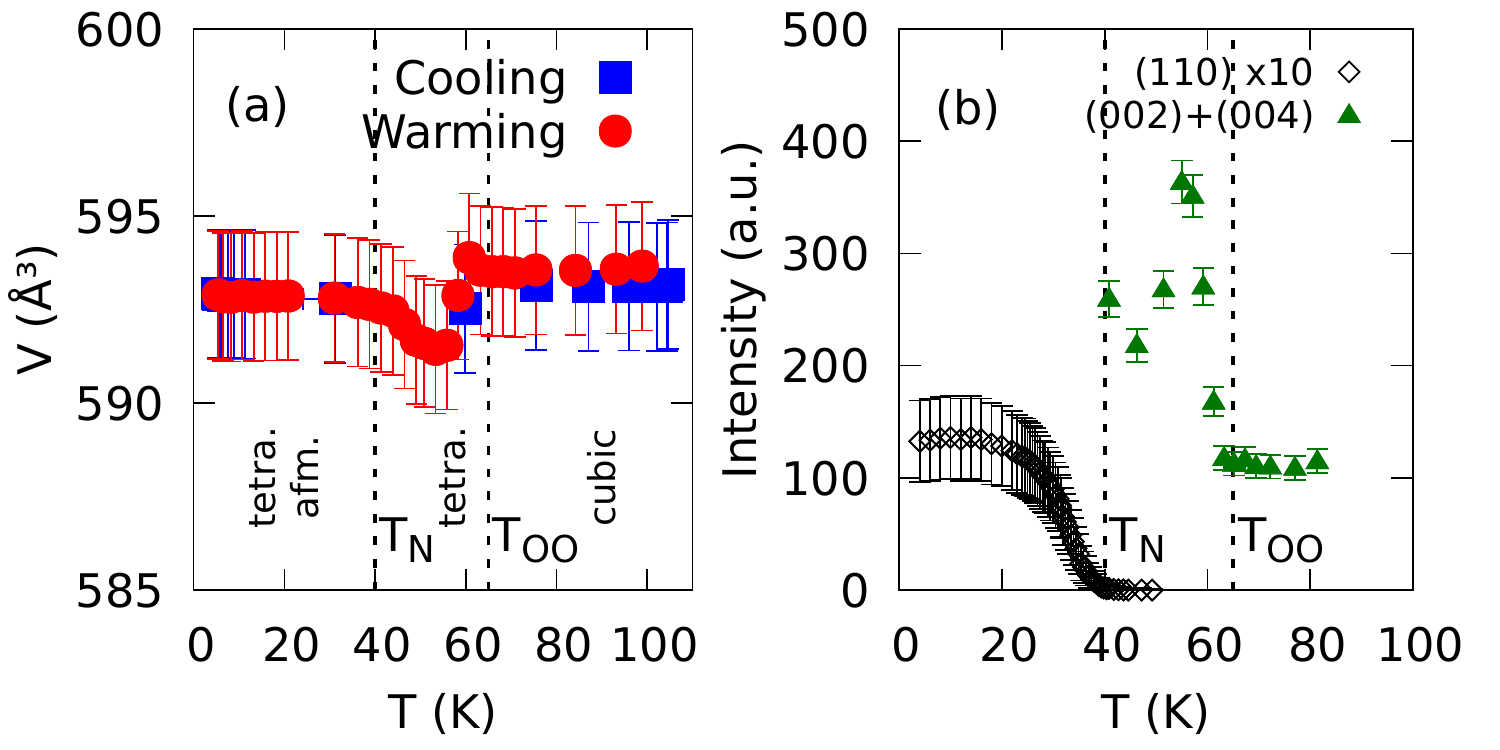}
	\caption{ \label{fig:lattice}$(a)$ Temperature-dependent change of the unit cell
	volume $V$ as determined using neutron diffraction.
	Note that the error bars include the instrument resolution as systematic error.
	$(b)$ The intensities of the $\vec{Q}$=(110) and the $\vec{Q}$=(004) Bragg reflections
	as a function of temperature illustrating the structural transition at T$_{OO}$=63 K
	and magnetic ordering temperature of T$_{N}$=42 K.
	The data in panel $(b)$ was taken using the RITA spectrometer in two-axis mode with E$_{i}$=5 meV.
	The (004) Bragg peak was measured at the (002) position using $\lambda/2$ with the scattered Be
	filter removed. Note that the intensity of (110) has been magnified by a factor of 10 relative to
	the (004) data.}
\end{figure}

At the neutron diffractometer MIRA~\cite{MIRA, MIRAnew}, the temperature ($T$) dependence
of the lattice constants was measured using $\theta$-$2\theta$ scans around
the (220) Bragg reflections of the single-crystals. The volume of the unit cell
was determined to be $V \approx 593\,\angs^3$. The very weak $T$-dependence of
$V$ (Fig. \ref{fig:lattice}) is compatible with the results of Refs.
\onlinecite{Wheeler10, Mamiya97}.

\subsection{ \label{sec:energies} Phonon dispersion}

We first describe the softening in energy of the acoustic fluctuations in \mvo{}.
Inelastic neutron scattering experiments were performed using the triple-axis
spectrometer (TAS) EIGER~\cite{Eiger2017} at the spallation source SINQ~\cite{Fischer97}
at the Paul Scherrer Institute (PSI) in Villigen, Switzerland
and the TAS IN22~\cite{Regnault1999} at the Institute Laue-Langevin (ILL) in
Grenoble, France.
Both spectrometers were operated in the constant final energy mode with
$E_f = 14.68\,\mathrm{meV}$. At EIGER and IN22, the dispersion of the acoustic
phonon branches was measured using a vertically focusing monochromator and
horizontally focusing analyser without using any collimation. For the EIGER measurements,
crystals oriented in the $hk0$ and the
$hhl$ planes were used, collecting data around the $(400)$ and $(440)$ Bragg
peaks. For IN22 we used the $hk0$ scattering plane.

In addition, experiments at small reduced momentum transfer $q$ were conducted
using the cold TAS option of MIRA~\cite{MIRA, MIRAnew} at the MLZ in Garching, Germany,
using neutrons with fixed incident energies
$3.78\,\mathrm{meV} < E_i < 4.97\,\mathrm{meV}$ yielding an energy resolution
(full-width at half-maximum, FWHM) of $0.11\,\mathrm{meV} < \Delta E < 0.17\,\mathrm{meV}$.
At MIRA, a vertically focusing monochromator was combined with a flat analyser.
No collimation was installed in the incident and scattered beams. Higher order neutrons were
removed by means of a cooled beryllium filter. At MIRA, we used the $hk0$ scattering plane and
collected data in the $(220)$ Brillouin zone.

Furthermore, we employed the time-of-flight spectrometer (TOF) MERLIN~\cite{MERLIN}
at the Rutherford Appleton Laboratory in Harwell, UK. MERLIN uses
a multi repetition-rate chopper, where we selected incident energies of
$E_i = 24\,\mathrm{meV}$ and $E_i = 49\,\mathrm{meV}$ and used a Fermi chopper
frequency of 350 Hz. Data was collected with the sample rotation covering the
full range of $90^{\circ}$ around the $(440)$ Bragg reflection and the
crystal oriented in the $hhl$ plane. Analysis was performed using the software
HORACE~\cite{Horace}.

The primary goal of our experiments was to establish the dispersion of the
TA1, TA2, and LA phonon modes.
Note that in our notation, the elastic constant $C_{44}$ is related to the sound
velocity of the twofold degenerate transverse mode TA1 propagating along a
$[001]$-direction with a polarisation along $[100]$ or $[010]$. $\left(C_{11}-C_{12}\right)/2$
corresponds to the TA2-mode propagating along $[110]$ with a polarisation along
$[1\overline{1}0]$. $\left(C_{11}+C_{12}+2C_{44}\right)/2$ is the sound velocity
of the LA-phonon propagating along $[110]$.  Owing to the neutron cross section
and selection rules associated with phonon eigenvectors~\cite{Harada70}, sensitivity
to both the TA1 and TA2 phonon modes is obtained with triple-axis measurements when
the crystal is aligned in the $hk0$ scattering plane while alignment in the $hhl$ plane
only affords measurements of the TA1 phonon.
Time of flight measurements using chopper spectrometers allow momentum transfers out
of the horizontal scattering plane to be measured and therefore afford sensitivity
to both TA1 and TA2.

\begin{figure}[htb]
	\includegraphics[width=1\columnwidth]{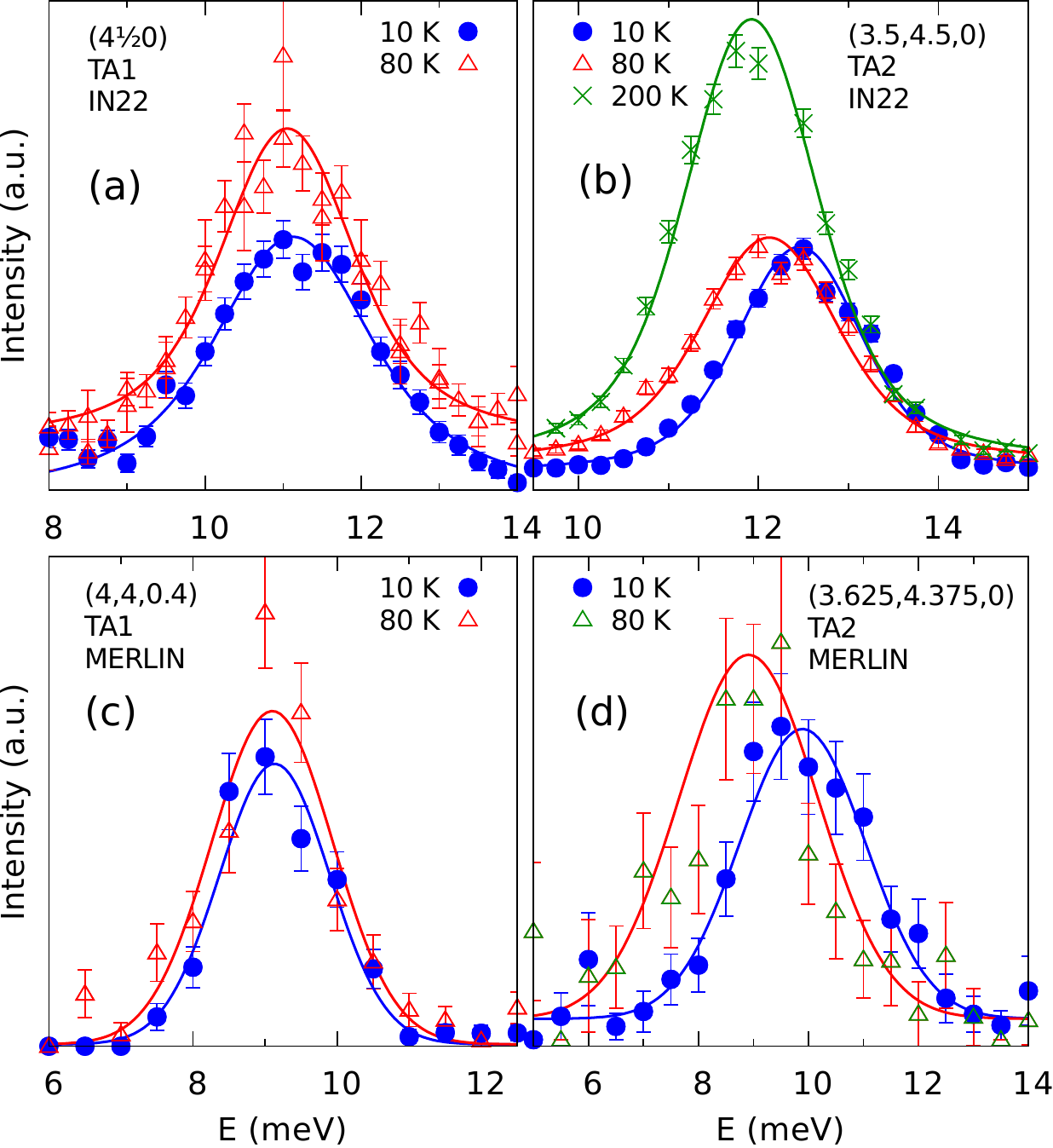}
	\caption{\label{fig:scans}
	Typical data for transverse phonons TA1 and TA2 are shown for $T = 10$ K and
	$T = 80$ K. There is a clear softening of the TA2 mode in the cubic phase
	when compared with the low temperature tetragonal phase.
}
\end{figure}

\begin{figure*}[htb]
	\includegraphics[width=1\textwidth]{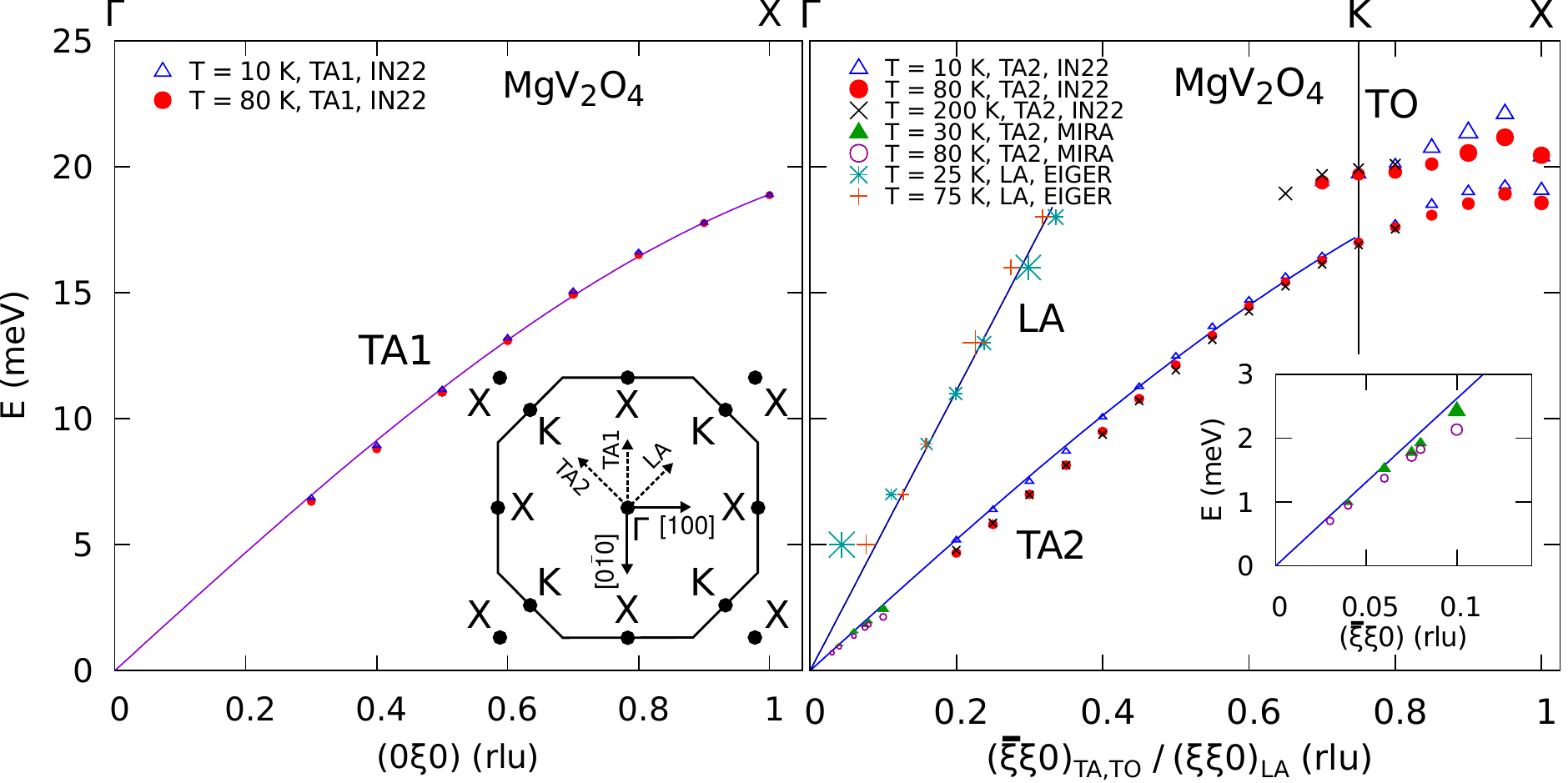}
	\caption{\label{fig:energies}
		The dispersion of the TA1 and the LA modes is independent of $T$ when
		going from the tetragonal to the cubic phase. When comparing the cubic
		($T > 63$ K) to the tetragonal ($T < 63$ K) phonon branches, the TA2
		modes show a ``spoon-like'' behaviour at intermediate $q$ values. The
		sizes of the symbols indicate the sizes of the error bars. The inset in
		the left panel shows the $hk0$ plane of the face-centred cubic Brillouin
		zone and the scan directions. The inset in the right panel shows a
		magnification of the low-$q$ TA2 phonon branch.
	}
\end{figure*}

Typical data of TA1 and TA2 phonons are shown in Fig. \ref{fig:scans} for
temperatures $T = 10$ K (tetragonal phase) and $T = 80$ K (cubic phase). In
panels (a) and (b) example $(4q0)$ (TA1) and $(4-q, 4+q, 0)$ (TA2) phonon data
from IN22 are shown. Panels (c) and (d) depict typical $(44q)$ (TA1)
and $(4-q, 4+q, 0)$ (TA2) slices from MERLIN. The TA1 mode does
not show any $T$-dependence while the TA2 mode becomes softer in energy
in the high temperature cubic phase.
We selected the example $q$ values shown in Fig. \ref{fig:scans} as they
are in the region where the effect is strongest and where the peaks can be
clearly separated from other contributions, e.g. incoherent scattering.
Fig. \ref{fig:energies} summarises the results of all phonon measurements that
were conducted at various temperatures in the range $10$ K $\le T \le 200$ K.
A transverse optic phonon branch is visible at $E \approx 20\,\mathrm{meV}$
near the zone boundary.
Attempts using both triple-axis and time of flight measurements failed to
track this mode closer to the zone centre.

The dispersion of the LA and TA1 phonon branches do not show appreciable changes
between the cubic and the tetragonal phases. In contrast, the TA2 branch
exhibits an observable $T$-dependence. At intermediate $q$-values, the
dispersion shows a ``spoon-like'' anomalous behaviour, i.e. the phonons soften in energy
when entering the cubic phase for $T > T_{OO}$.
While the data is suggestive of a softening in energy of the TA2 phonon near the zone centre,
the effect is not as large as at intermediate wavevectors.
The effect at small $q$ close to the zone centre $\Gamma$ is also less pronounced than the
$q \rightarrow 0$ behaviour reported by Watanabe \textit{et al.}
\cite{Watanabe14} using ultrasound sensitive to acoustic fluctuations on the MHz timescale.

\begin{figure}[htb]
	\includegraphics[trim=5bp 2bp 10bp 2bp, clip, width=1\columnwidth]{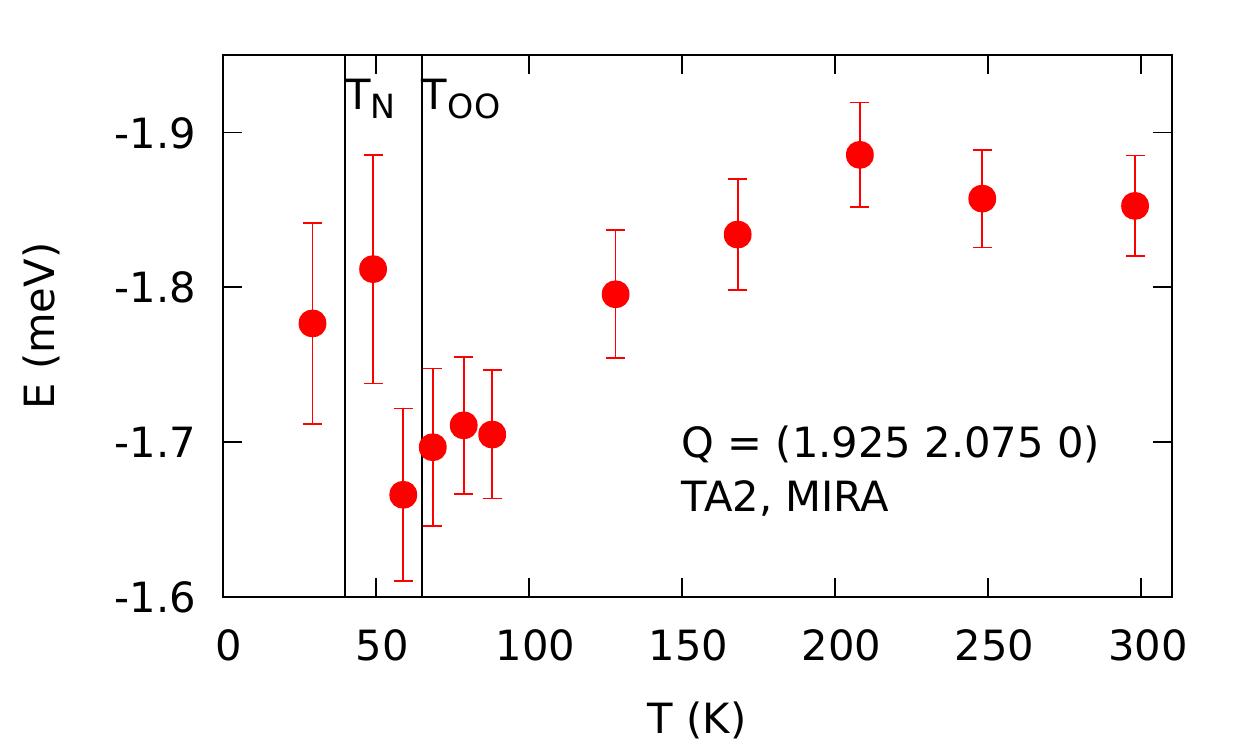}
	\includegraphics[trim=5bp 2bp 10bp 2bp, clip, width=1\columnwidth]{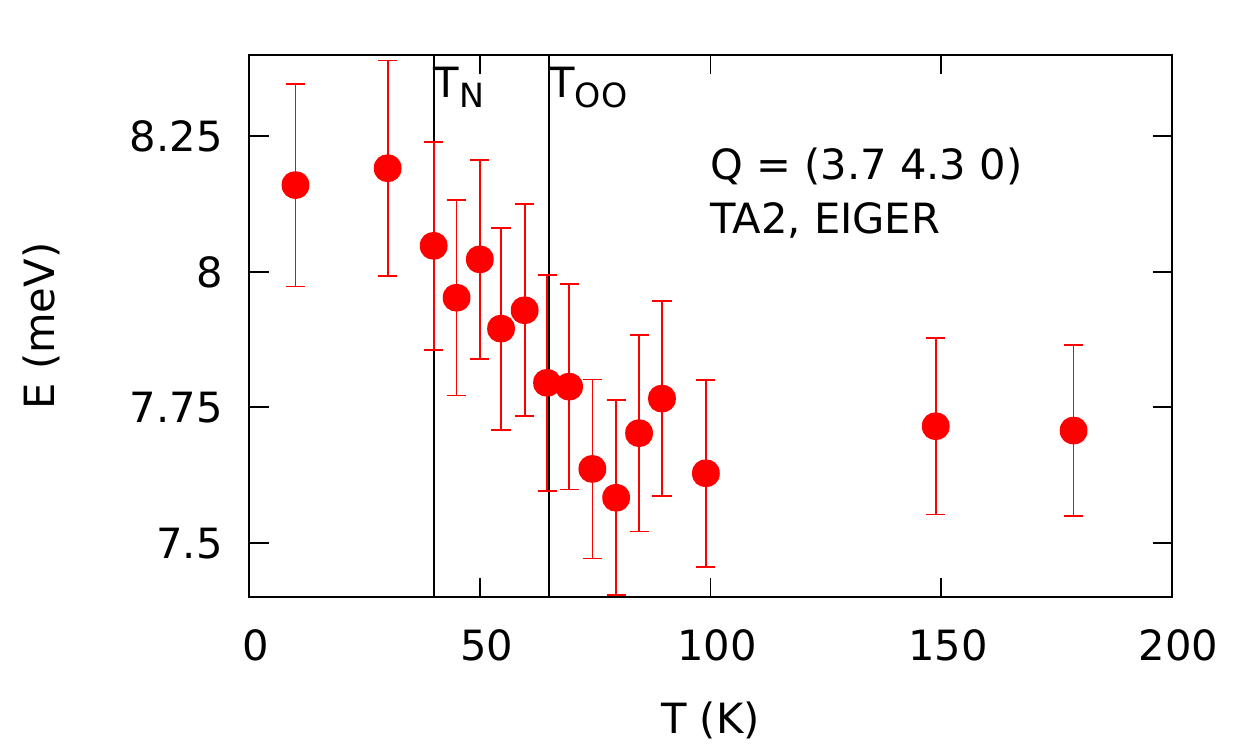}
	\includegraphics[trim=5bp 2bp 10bp 2bp, clip, width=1\columnwidth]{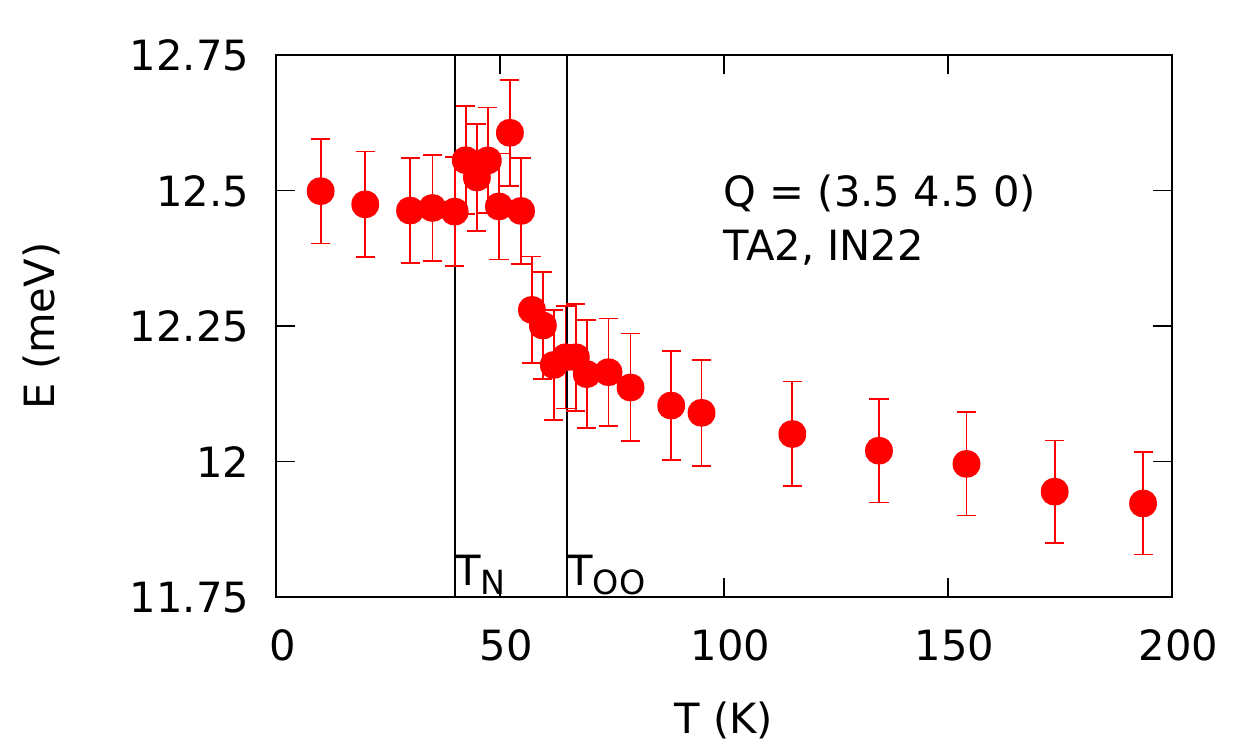}
	\caption{\label{fig:detailscan}
		The dispersion of the TA2 phonons softens and hardens at
		$q = (\overline{0.075}\ 0.075\ 0)$ (top) and
		$q = (\overline{0.5}\ 0.5\ 0)$ (bottom),
		respectively, when approaching the cubic-tetragonal phase transition
		near $T_{OO} = 63$ K from high temperatures. For
		$q = (\overline{0.3}\ 0.3\ 0)$ (middle) there is no change in phonon
		energies in the $T > T_{OO}$ region. Please note that at MIRA we
		measured using neutron energy gain ($-E$) for reasons of
		analyser efficiency and angular constraints at the instrument. The
		physics of the system is unaffected by this choice.
	}
\end{figure}

The detailed  temperature dependence of two TA2 phonons at
$q = \left(\overline{0.075}\ 0.075\ 0\right)$,
$q = \left(\overline{0.3}\ 0.3\ 0\right)$, and
$q=\left( \overline{\frac{1}{2}} \frac{1}{2} 0 \right)$ are shown in Fig.
\ref{fig:detailscan}. Both the modes at small and large $q$
soften by ca. 0.1 meV and 0.4 meV, respectively, when entering the cubic phase.
For $T_{OO} < T < 200 $ K, the low-$q$ phonons harden by ca 0.2 meV, while the
large-$q$ modes further soften by ca. 0.2 meV.

\subsection{ \label{sec:linewidths} Phonon linewidths }

Having established the presence the softening over intermediate wavevectors of
the TA2 acoustic phonon, the linewidth indicative of the phonon lifetime is
now discussed.  We have determined the linewidth of phonons by analysing the inelastic
scattering data from EIGER and IN22 using the dynamical structure factor given
by Ref. \onlinecite{Fak1997} which obeys detailed balance required for neutron cross sections
\begin{multline} \label{eq:Sqw}
  S\left(q,E\right) = \frac{S_0}{\left|A\cdot\sin\left(x\cdot q\right)\right|
	\cdot\pi} \cdot \left[ \frac{1}{\exp \left(E / k_BT\right) - 1} + 1 \right] \cdot \\
	\left(\frac{\pm\Gamma_p}
		{\left(E\mp \left|A\cdot\sin\left(x\cdot q\right)\right| \right)^{2}+\Gamma_p^{2}}\right).
\end{multline}

Here, the dispersion of the TA2 branch is modeled with a simple sine function
that is appropriate to describe acoustic phonons. The lineshape of the phonons
is approximated with a Lorentzian function. The
reduced momentum transfer is given by $q = 2\pi\sqrt{(h^2 + k^2)/a^2 + l^2/c^2}$,
where $a$ and $c$ are the lattice constant defining the tetragonal unit cell.
The term in square brackets is the Bose occupation factor for temperature $T$
and energy $E$. $\Gamma_p$ denotes the half-width at half-maximum of the phonon,
$S_0$ is an overall scaling factor. For $\Gamma < q < \mathrm{K}$ the TA2
dispersion at low temperature (Fig. \ref{fig:energies}) can be approximated by
$A = \left( 22.5 \pm 0.1 \right) \, \mathrm{meV}$ and
$x = \left(0.825 \pm 0.004 \right)$. Here, $A$ and $x$ serve as scaling factors
for the energy and reduced momentum of the dispersion, respectively.

$S(q,E)$ as given by Eq. (\ref{eq:Sqw}) was convoluted with the
Eckold-Sobolev resolution function~\cite{Eckold2014} of a TAS spectrometer using
Monte-Carlo integration. This novel algorithm was used because it reproduces the
resolution function of focusing TAS better than the algorithm of
Ref. \onlinecite{Popovici75}. Fitting the data was performed using Minuit's
simplex minimiser~\cite{Root2011}. A full description of the software tool
\textit{Takin} that was developed by some of the authors and used for the
convolution fits can be found in Refs. \onlinecite{Takin16, Takin17}.

\begin{figure}[htb]
	\includegraphics[trim=5bp 5bp 5bp 5bp, clip, width=1\columnwidth]{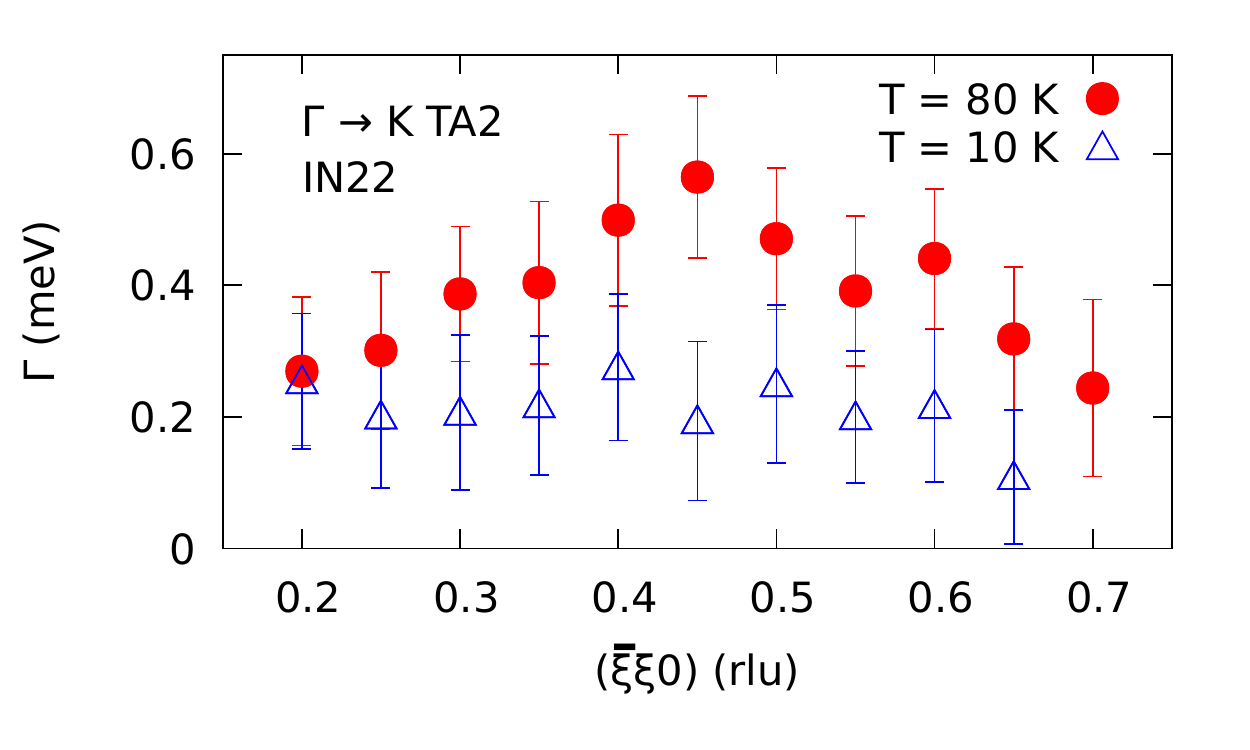}
	\caption{\label{fig:linewidths_q} Linewidth $\Gamma_p$ of the TA2 phonon
	branch. In the tetragonal phase at 10 K (blue triangles), $\Gamma_p$ is
	independent of momentum transfer. In the cubic phase at $T = 80$ K (red
	points), $\Gamma_p$ attains a maximum halfway from the $\Gamma$ to the
	$\textrm{K}$ point.}
\end{figure}

The $q$-dependence of $\Gamma_p$ of the TA2 branch is shown in Fig.
\ref{fig:linewidths_q}. While $\Gamma_p$ is essentially $p$ independent at 10 K,
it attains a maximum about half-way from the $\Gamma$-point to the K-point at
80 K. The maximum occurs at the $q$-position where we also observe the strongest
softening of the TA2 phonon branch (Fig. \ref{fig:energies}). No appreciable
changes were observed for the other acoustic phonon branches or for the linewidth
of the TA2 phonons in the tetragonal phase.

\begin{figure}[htb]
	\includegraphics[trim=5bp 5bp 5bp 5bp, clip, width=1\columnwidth]{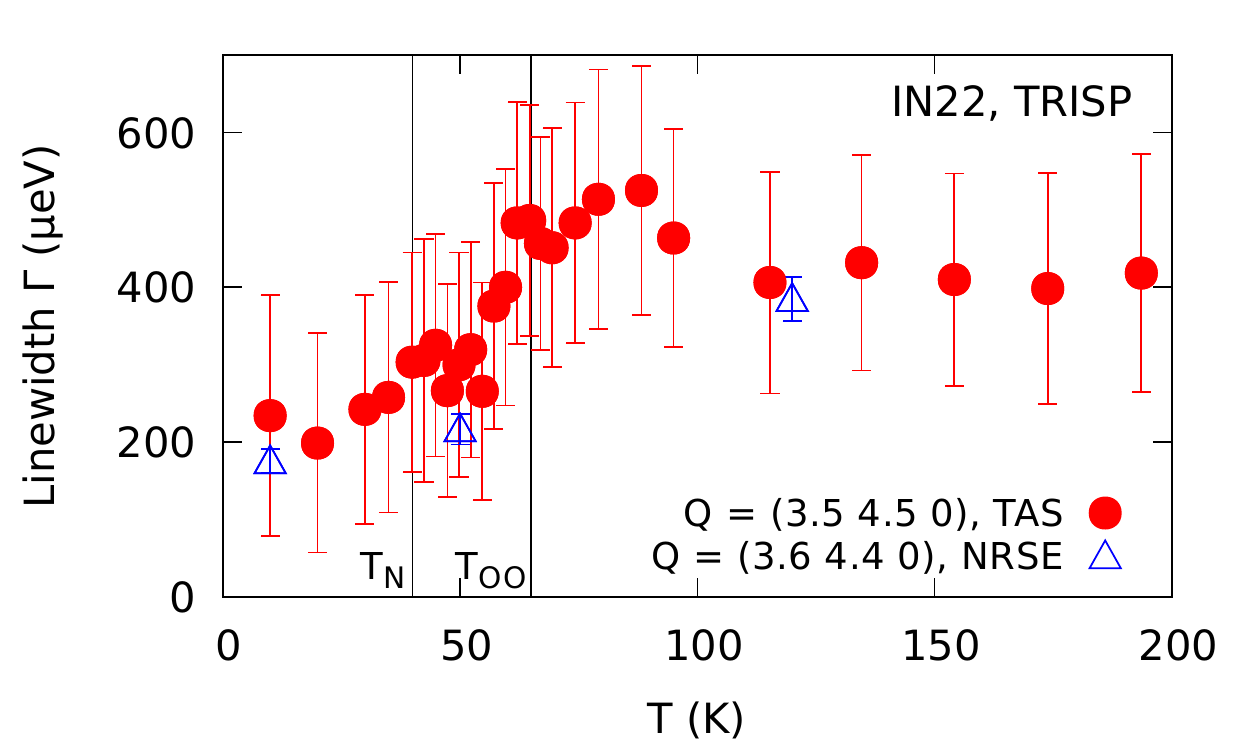}
	\caption{\label{fig:linewidths_T} Temperature dependence of the TA2 phonons
	at $Q = (3.5\ 4.5\ 0)$ using conventional TAS-spectroscopy (red dots) and
	$Q = (3.6\ 4.4\ 0)$ using TRISP (blue triangles). A subtle plateau can be
	identified between $T_N$ and $T_{OO}$. $\Gamma_p$ increases to approximately
	0.45 meV in the cubic phase where it remains constant up to $T = 200$ K.}
\end{figure}

The temperature dependence of the TA2 phonons at
$q = (\overline{\frac{1}{2}} \frac{1}{2} 0)$ and
$q = (\overline{0.4}\ 0.4\ 0)$ are shown in Fig. \ref{fig:linewidths_T}.
$\Gamma_p$ increases when warming the sample from $T = 10\,\mathrm{K}$ towards
the N\'eel-transition at $T_{N} \approx 40\,\mathrm{K}$. Here $\Gamma_p$ seems
to saturate at $\approx 200\, \mathrm{\upmu eV}$ before increasing again to a
maximum value $\Gamma_p \approx 450\, \mathrm{\upmu eV}$ in the vicinity of the
structural phase transition $T_{OO} \approx 63\,\mathrm{K}$.

Note that the linewidth at $q = (\overline{0.4}\ 0.4\ 0)$ was determined with
high accuracy by means of the neutron-resonance spin-echo (NRSE) technique at
the triple axis spectrometer TRISP at MLZ~\cite{TRISP}. TRISP was set up in a
negative-negative-positive scattering configuration. The NRSE coils were aligned
such that the focusing condition for the TA2 branch was fulfilled. The data
was corrected for resolution effects reducing polarisation~\cite{Habicht03, Habicht04}.
These are caused by the sample mosaic and the
slope of the phonon dispersion.

\subsection{ \label{sec:fields} Field dependence of phonon energy }

We finally present the magnetic field dependence of the transverse acoustic phonon
lifetimes and energies in the same wavevector region where temperature dependent anomalies are observed.
While a magnetic field response has been reported for polar phonons in spinels based
on optical data~\cite{RudolfA07}, magnetic field effects on the acoustic phonon response
have been reported in superconductors and also in metals where the Fermi surface topology
is relatively flat in momentum space~\cite{Pynn74}.
Attempts on semiconducting materials have failed to observe any effect.~\cite{Comes81}

\begin{figure}[htb]
	\includegraphics[trim=5bp 2bp 5bp 5bp, clip, width=1\columnwidth]{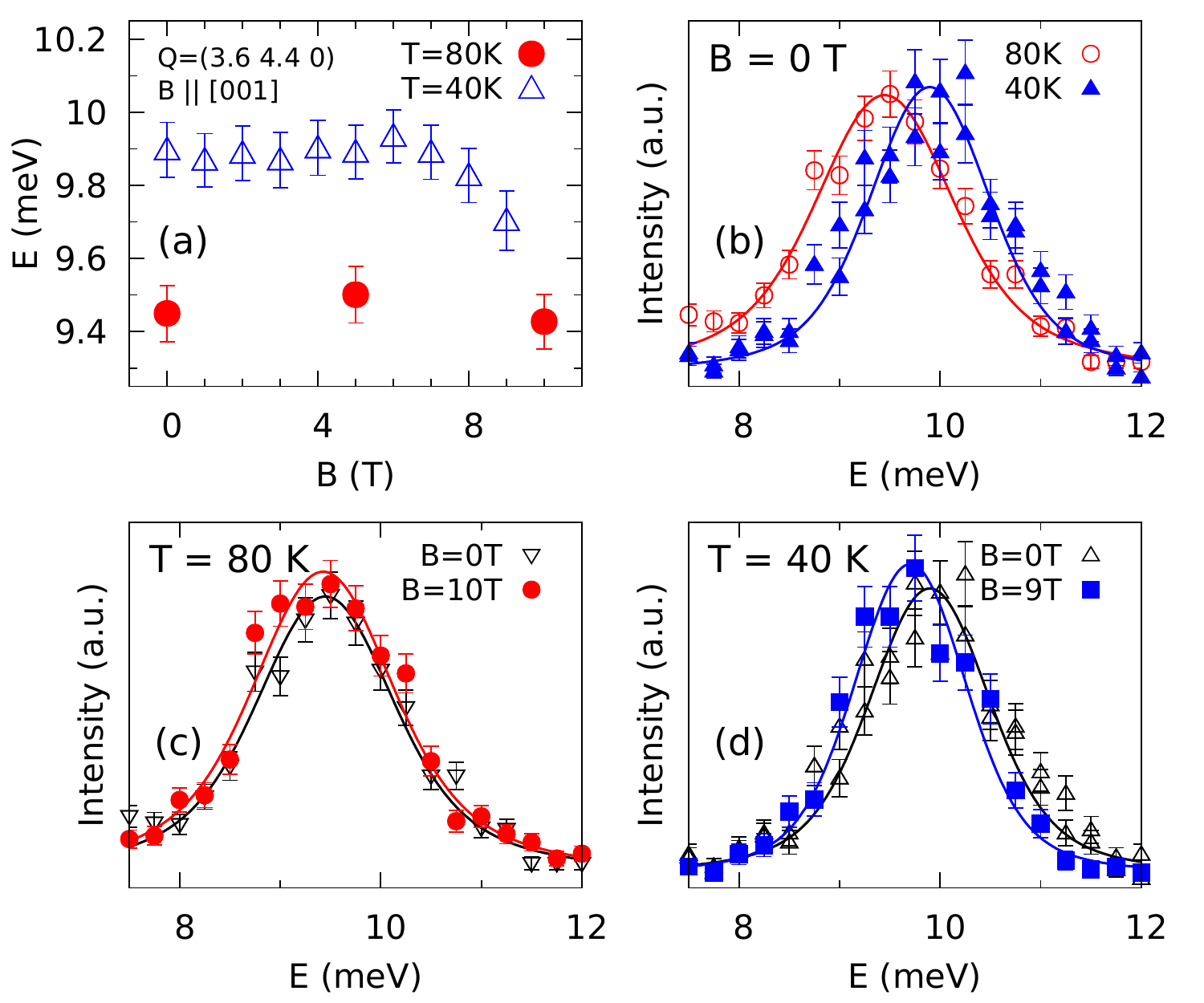}
	\caption{\label{fig:field_001} Panel $(a)$ shows the field-dependence of the TA2 phonon at the
	$q$-position $\left(\overline{0.4}\ 0.4\ 0\right)$. Panels $(b)$--$(d)$ depict selected phonon data.
	A small softening of $\approx 0.2$ meV is observed at high fields in the tetragonal phase for
	$T = 40$ K as shown by blue triangles in panel $(a)$ and in panel $(d)$.}
\end{figure}

The field dependence of $\Gamma_p$ of the TA2 phonon at  $q=\left(\overline{0.4}\ 0.4\ 0 \right)$
was determined for \mvo{} by application of a magnetic field $B_{[001]}$ along
the $[001]$-direction  (Fig. \ref{fig:field_001}).
The phonon energy $E$ is independent of $B_{[001]}$ in the cubic phase while $E$ decreases
subtly at $T = 40$ K in the tetragonal phase above 7 T by $\approx 0.2$ meV up to the highest
magnetic field achievable in this experiment.
The results are suggestive of a slight decrease in energy of the acoustic phonon at high
magnetic fields at intermediate wavevectors.
Further measurements to higher fields and also in other orbitally degenerate spinels
would be helpful for establishing this effect.

\section{ \label{sec:disc} Discussion }

By measuring the phonon dispersion along the high-symmetry directions $[0\xi0]$
and $[\overline\xi\xi0]$ using neutron scattering, we have demonstrated an anomaly
in the TA2 branch along the $\langle\xi\xi0\rangle$-directions associated with the
structural and orbital phase transition in \mvo{} at $T_{OO}$ (Fig. \ref{fig:energies}).
An energy softening at intermediate wavevectors mid-way between the $\Gamma$- and the
K-point is observed with increasing temperature.
This softening leads to a spoon-like dispersion of the TA2 branch.

We first consider the possibility of thermal expansion~\cite{Maradudin62} of the lattice
as an explanation for the acoustic phonon anomalies in \mvo{}.
A change in the unit cell volume would result in a change in the size of the Brillouin
zone and, assuming a fixed zone boundary energy, would give a corresponding shift in
phonon velocity.
This cannot explain our data for two reasons.
First, the phonon anomaly is only observable in \mvo{} over a range of intermediate
wavevectors and no change in the $lim_{q\rightarrow 0}$ slope is observed using neutron
scattering on the THz timescale.
Second, the change in phonon energy is larger then would be expected from such an effect
and is inconsistent with the thermal expansion data presented in Fig. \ref{fig:lattice}
and also from the higher resolution x-ray data for the unit cell volume presented in Ref. \onlinecite{Mamiya97,Kaur13}.
Therefore, the anomaly observed here is not due to thermal changes in the lattice constants.

The phonon anomaly in \mvo{} measured with neutron scattering differs from the response
observed with ultrasound~\cite{Ishikawa03}.
At $T_{OO}$, Watanabe \textit{et al.} \cite{Watanabe14} reported a temperature dependent
change of the sound velocity of the TA2 mode $\Delta v \approx 25\,\%$.
We find no observable change for momentum transfers near the zone centre, however
measure a spoon-like anomaly for intermediate wavevectors.
We speculate that the difference between the neutron low-$q$ data and the ultrasound
data in the $lim_{q \rightarrow 0}$ originates from the differing energy and
lengthscales of the two measurements.
Owing to temporal resolution, neutron scattering measure fluctuations on the THz
timescale while ultrasound measures the dynamic response for long wavelengths typically
on the MHz timescale.
Similar differences between neutron scattering and ultrasound have previously been
reported for spinels in Ref. \onlinecite{Thompson78}.
The neutron scattering results do not reflect an observable softening of the TA2
phonon on the THz timescale, rather an anomaly over a range of intermediate wavevectors
of $\Delta v \approx \left(12 \pm 5\right) \,\%$.

It is interesting to compare our results with theoretical predictions of soft acoustic
phonons near a Jahn-Teller distortion.
Considering the Jahn-Teller effect and coupling to acoustic phonons, a theoretical
model for the phase transition from the tetragonal to the cubic phase is given in papers by
Pytte~\cite{Pytte71, Pytte73}.
In the first paper, Pytte presents a theory which models the temperature dependence of
the elastic constants as a result of the coupling between the elastic strain tensor and
the twofold degenerate $d$ orbitals.
Including higher-order anharmonic coupling, the  $T$-dependence of the transverse
acoustic phonon mode propagating along $[110]$ with polarisation along
$[1\overline{1}0]$ -- the tetragonal shear elastic mode -- shows a softening with
decreasing temperature, a step discontinuity at the transition temperature $T_{OO}$,
followed by an increase of the velocity of sound for further decreasing temperature.
The second paper~\cite{Pytte73} includes the coupling of the degenerate orbitals to
the optical phonon modes and finds a ``central peak'' in the dynamical structure factor
$S\left(q,\omega\right)$. Central peaks have been reported to often co-occur with soft
modes and had first been identified in $\mathrm{SrTiO_3}$ \cite{Riste71} and
$\mathrm{Nb_3Sb}$ \cite{Shirane71,Axe73}.
They are also discussed theoretically in Refs. \onlinecite{Halperin76, Cowley06}.

The theory developed by Pytte predicts temperature dependent changes to the TA2 phonons
(with velocity proportional to $C_{11}-C_{12}$) and this is the same mode which displays
anomalies in \mvo{} over intermediate wavevectors.
However, there are several important differences which make the relevance of this theory
to the case here questionable.
The theoretical prediction is in the context of a two-fold orbital degeneracy associated
with the partial occupancy of the $e_{g}$ orbitals.
As outlined above, this is not the case for S=1 V$^{3+}$ where two electrons occupy
the $t_{2g}$ orbitals resulting in a three-fold orbital degeneracy.
Pytte~\cite{Pytte71} also predict a true soft mode whose energy reaches zero at the phase
transition for momenta $q \rightarrow 0$.
A general theory for transitions driven by orbital degeneracy by Elliott~\cite{Elliott1977}
and Young~\cite{Young75} also reach the same conclusion with regards to the prediction of
a soft mode close the zone centre.
Experimentally, this would be reflected in the neutron response by a softening of the TA2
acoustic phonon branch and is more compatible with the results from ultrasound and also in
agreement with neutron scattering experiments studying the low energy acoustic phonons in
Jahn-Teller active PrAl$_{3}$~\cite{Birgeneau74}.
In contrast to these predictions and previous experimental examples for Jahn-Teller driven
structural distortions, we observe an anomaly at finite-$q$ over a limited intermediate range
in wavevector.
Another discrepancy concerns the central peak at $E=0$ \cite{Pytte73}, which we do not observe
at finite $q$.
The strong increase in zone centre Bragg scattering upon cooling into the tetragonal phase,
which had previously been reported by Wheeler \textit{et al.} \cite{Wheeler10}, may be instead
related to the central peak, however it appears at a different $q$ than the acoustic phonon
anomaly. Based on these comparisons with theory and experiment, it is difficult to associate
the TA2 phonon anomaly in \mvo{} with a soft mode due to a proximate Jahn Teller distortion
resulting in breaking of the ground state orbital degeneracy and a soft acoustic mode in
the $lim_{q\rightarrow 0}$.

The simultaneous broadening and softening of an acoustic phonon over a range in wavevector
transfers is indicative of coupling to some other degree of freedom (see for example discussion
of coupling in Ref. \onlinecite{Wehner75,Waki02,Hlinka03,Stassis97}).
Coupling of acoustic phonons to crystal field~\cite{Bruhl78} driven distortions have been
experimentally studied with examples being the TA2 phonon anomalies reported in DyVO$_{4}$ and
TbVO$_{4}$.~\cite{Melcher72,Sandercock72}
Similar coupling effects have also been discussed in UO$_{2}$.~\cite{Allen68}
These examples involve the coupling of an acoustic phonon to a crystal field level with a similar
energy scale to that of the phonon and this is also predicted to be required in the case of
magnetic pyrochlores.~\cite{Yamashita00}
While the $dd$ transitions in $3d$ transition metal ions are large
($\sim$ 1-2 eV)~\cite{Haverkort07,Kant08,Schooneveld12,Kim11,Cowley13} making them unlikely to
be involved with this process, the lower energy spin-orbit split levels may have a more
appropriate energy scale.
As discussed by Tchernyshyov~\cite{Tcherny04}, the ground state of V$^{3+}$ derived by
diagonalising the spin-orbit Hamiltonian is a $j_{eff}$=2 quintuplet which is separated by an
energy scale of $\sim$10 meV~\cite{Abragam:book} to an excited $j_{eff}$=1 triplet state.
The splitting of such degenerate ground state due to a local molecular field could provide the
spin-orbit crystal field level with the correct energy scale to couple to the acoustic phonon.
However, we emphasise that such a coupling in the framework of the theories described above
would involve a true softening of the fluctuations near the Brillouin zone centre inconsistent
with the phonon anomalies in \mvo{}.

Another possibility is coupling between the acoustic phonons and low energy magnon excitations
from the V$^{3+}$ S=1 sites.
No magnon excitations were observed to cross the phonon branch in our experiments, however, it
should be noted that the comparatively large momentum transfers optimised for phonon measurements
are where magnetic form factors ensure reduced neutron scattering cross sections from magnons.
But, such magnetic excitations have been reported to be highly dispersive~\cite{Wheeler10}, gapped~\cite{Gleason2014},
and located in reciprocal space away from the low-energy acoustic lattice fluctuations, making
coupling in the context of the theories discussed above likely weak.
While full magnon dispersion curves from neutron scattering are limited in \mvo{}, neutron
measurements in ZnV$_{2}$O$_{4}$~\cite{Lee04} observe a significant gap of $\sim$ 10 meV at
the magnetic zone centre while we observe anomalies at a displaced momentum and at lower energies.
Based on the energy scale in comparison to analogous systems and the different momentum transfer
away from either the zone boundary or centre, we conclude that coupling to magnons is unlikely.

It is interesting to compare the results in \mvo{} to metallic systems with acoustic instabilities.
Our results show similarities to the measurements by F. Weber \textit{et al.}  \cite{Weber2011}
who determined the dispersion of the TA-phonons in a metallic manganite.
In this material, a charge and orbital ordering (COO) sets in at low temperatures.
Above $T_{COO}$, F. Weber \textit{et al.} \cite{Weber2011} report a softening of the
TA phonon at $q = \left( \frac{1}{4} \overline{\frac{1}{4}} 0 \right)$ by $\Delta E \approx 0.25 \  \mathrm{meV}$
accompanied by an increase in linewidth  $\Delta \Gamma_{HWHM} \approx 0.25 \,\mathrm{meV}$.
Similar anomalies have been reported in other metallic manganites on decreasing temperature.~\cite{Hoesch2013}
The anomaly in these example manganites is associated with electron-phonon coupling for wave
vectors associated with Fermi surface nesting, given the metallic electronic response.
Indeed, similar intermediate acoustic phonon anomalies have been reported in a number of
materials where there is a coupling between lattice and charge degrees of freedom.
One example is the one-dimensional conductor TTF-TCNQ~\cite{Shirane76} which displays a
softening of a longitudinal acoustic phonon over intermediate wavevectors near a Peierls
transition and also in superconductors where the change in linewidth has been related to
the onset of a gap in the electronic quasiparticle response~\cite{Shapiro75,Weber2008,Weber2014,Keller06,Bullock57}.
The wave vector at which the acoustic mode is unstable in these examples is determined by
the electronic Fermi surface and hence the metallic properties.

While the analogy between soft acoustic modes in metallic systems is not obvious given that
\mvo{} is an insulator, there are several properties that indicate that charge fluctuations may
be playing a role in this spinel.
There is evidence that \mvo{}, ZnV$_{2}$O$_{4}$, and CoV$_{2}$O$_{4}$ are proximate to an
insulator-metallic transition under pressure evidenced through both calculations~\cite{Canosa07}
and also thermodynamic and transport measurements~\cite{Kism11}.
This has led to theoretical studies indicating that both \mvo{} and ZnV$_{2}$O$_{4}$ contain
large charge fluctuations and should be considered in a partially delocalised regime.~\cite{Kato12}
Therefore, both theory and experiment are suggestive of electronic or charge fluctuations
in \mvo{} and the possible coupling between the acoustic phonons and the charge channel is
further corroborated by our magnetic field measurements which would alter the chemical
potential.~\cite{Maitra07}
While \mvo{} is an insulator, the similarity between the acoustic phonon anomaly presented above
and the studies in metallic systems discussed here support the notion of charge fluctuations
in \mvo{}.

The similarity between the \mvo{} phonon anomalies and metallic systems discussed above is
reminiscent of the Peierls transition in one dimensional systems, with TTF-TCNQ discussed above
being an example.
This analogy may be appropriate for \mvo{}~\cite{Tcherny02} given the one dimensional
magnetic V$^{3+}$ chain interactions which have been measured using neutron
spectroscopy~\cite{Wheeler10}.
Similar one dimensional magnetic correlations have been reported in ZnV$_{2}$O$_{4}$~\cite{Lee04}
and also in MgTi$_{2}$O$_{4}$~\cite{Schmidt04}, both of which have a triply orbital degenerate
ground state.
Supporting this possibility further, spin dimerisation has been reported in the spinel
CuIr$_{2}$S$_{4}$~\cite{Radaelli2002} where both Ir$^{3+}$ and Ir$^{4+}$ are present.
Given this similarity to other spinels and the one dimensional magnetic correlations,
an instability towards dimerisation resulting from an ``anti-Jahn Teller" distortions has
been predicted to apply to \mvo{} and other orbitally degenerate V-based
spinels,~\cite{Khomskii05} as well as the Verwey transition in Fe$_{3}$O$_{4}$~\cite{Shapiro75}.
The T$_{2}$ acoustic phonon anomalies we observe in \mvo{} are along the same direction as the
strong magnetic exchange resulting from orbital ordering resulting from the Jahn Teller
distortion, but we emphasise the softening and dampening is most prominent in the high
temperature cubic phase.
While the analogy with the Peierls transition in one dimensional systems might be compelling,
to our knowledge there is no report of structural dimerisation in either ZnV$_{2}$O$_{4}$ or \mvo{}.
We speculate that the acoustic phonon anomaly in \mvo{} may indicate the close proximity to
such a structural phase without the actual formation of long-range structural order.
It may even be indicative of spatially localised structural distortions that would
increase the acoustic phonon lifetime in a similar manner to that reported in disordered
piezoelectrics.~\cite{Stock12}
Supporting this, we note that additional structures are known to compete with the low
temperature tetragonal phase.~\cite{Suchomel12}

In conclusion, we identify a softening and damping of the TA2 phonon modes over a range
of intermediate wavevectors.
Neutron measurements of the acoustic phonons at low momentum transfers do not show any
observable softening on the THz timescale in contrast to expectations of a uniform
softening of the mode due to a change in the elastic constants.
The combined linewidth broadening and also softening indicates a coupling between the
acoustic TA2 phonon and another degree of freedom, analogous to electron-phonon coupling
observed in metallic compounds with a structural instability.
The results are suggestive of coupling to charge fluctuations predicted by theory.

\begin{acknowledgments}

We thank R. Schwikowski and A. Mantwill for technical support.
Financial support from the EPSRC is gratefully acknowledged.
This work is based on experiments performed at the Swiss spallation neutron
source SINQ, Paul Scherrer Institute (PSI), Villigen, Switzerland. The project
has received funding from the European Union's Seventh Framework Programme for
research, technological development and demonstration under the NMI3-II Grant
number 283883.
The authors gratefully acknowledge the financial support provided by JCNS to
perform the neutron scattering measurements at the Institute Laue-Langevin (ILL),
Grenoble, France.
This work is furthermore based upon experiments performed at the TRISP
instrument operated by MPG and the MIRA instrument at the Heinz Maier-Leibnitz
Zentrum (MLZ), Garching, Germany.
Experiments at the ISIS Pulsed Neutron and Muon Source were supported by a
beamtime allocation from the Science and Technology Facilities Council.
This work was part of the Ph.D. thesis of T. Weber \cite{PhDWeber}.
\end{acknowledgments}

\bibliographystyle{aipnum4-1}
\bibliography{mvo}

\end{document}